\documentclass{acmsmall}
\title{Topic words analysis based on LDA model}
\author{XI QIU
	\affil{Ohio State University}}
\begin{abstract}
Social network analysis (SNA), which is a research field describing and modeling the social connection of a certain group of people, is popular among network services. Our topic words analysis project is a SNA method to visualize the topic words among emails from Obama.com to accounts registered in Columbus, Ohio. Based on Latent Dirichlet Allocation (LDA) model, a popular topic model of SNA, our project characterizes the preference of senders for target group of receptors. Gibbs sampling is used to estimate topic and word distribution. Our training and testing data are emails from the carbon-free server Datagreening.com. We use parallel computing tool BashReduce for word processing and generate related words under each latent topic to discovers typical information of political news sending specially to local Columbus receptors. Running on two instances using paralleling tool BashReduce, our project contributes almost 30\% speedup processing the raw contents, comparing with processing contents on one instance locally. Also, the experimental result shows that the LDA model applied in our project provides precision rate 53.96\% higher than TF-IDF model finding target words, on the condition that appropriate size of topic words list is selected. 
\end{abstract}
\category{}{}{}
\terms{}
\acmformat{}
\begin{document}
\begin{bottomstuff}
Author's address: X. Qiu, Department of Computer Science and Engineering, Ohio State University, Columbus, OH 43210, USA.
\end{bottomstuff}
\maketitle

\section{Introduction}

Social network is a network consisting of constant information of
the connection of a group of people. Social Network Analysis (SNA)
\cite{Adamic04} discovers the unique or target characters to describe the
certain pattern of such people connection. As an implementation of
SNA, online social services like \textit{LinkedIn.com} are becoming
popular these years. In some previous studies, SNA models are implemented
to make connections of certain groups by directed graph models or
weighted edge graph models. For example, Wang and his coworkers develop
a probabilistic factor graph model \cite{Wang10} to analyze bibliographic
network in academic world. When conducting research in social networks
by characterizing documents, such as emails and academic publications,
data mining models have been applied in many SNA models. Lada Adamic
and Eytan Adar present Latent Dirichlet Allocation (LDA) model \cite{Blei03},
which selects Dirichlet distribution to estimate topic mixture, and
samples training data by Expectation Maximization (EM) algorithm.
This LDA model collects separated data to improve predicted models,
generating representative words under such latent topics from given
documents. Later, Rosen-Zvi and her group extend LDA model to Author
Topic (AT) model \cite{Zvi04}, determining topics by both content and
author distributions.

Our project collects topic words from emails and models
the preference of senders in typical website, using Gibbs sampling
as training strategy instead of EM algorithm. We create an email account
at Columbus, Ohio and use this account to register at \textit{Obama.com}.
Emails sent by \textit{Obama.com} are received through \textit{Datagreening.com}
\cite{Steward12}, which is a carbon-free server developed by our research
group. Then the downloaded emails are processed to input data for
LDA model in an appropriate format. Then we apply the LDA model mapping
the data to topic layers. Thus the topic words list is able to be
generated in the next step. Finally, by analyzing the similarity between
target words list and topic words list generated in above steps, we
establish which information are likely given to publicity by \textit{Obama.com}.

Python packages have been used to process words, such as
word tokenizing, word stemming and filtering out stop words, etc.
Word count program is run on two nodes implementing the parallel computing
tool BashReduce \cite{Zawodny09} to achieve almost 30\% speedup. TF-IDF model
\cite{Liu04} is selected to be a comparison generating topic words list.
The model applied in our project provides precision rate 53.96\% higher
than TF-IDF model, when we define the size of topic list properly.

The rest parts of this paper are organized in following
five sections: In section 2, we introduce the generative LDA model
and the design details of applying this model in our project. Section
3 describes parameter estimation using Gibbs sampling \cite{Gilks96}. Section
4 describes \textit{Datagreening.com} server, BashReduce tool and
python packages used in the project. Experimental results are shown
and evaluated in section 5. Conclusion and future work are discussed
in section 6.

\section{Generative LDA Model}

In this section, we will first briefly introduce the unigram model,
a model generating documents by single topics. Then we discuss LDA
model applied in our project. To simplify the process, we define that
a document is a collection of discrete words.

\subsection{Unigram Model}

Unigram Model, which is a basic probability model, treats
words as isolated elements in each document \cite{Nigam00}. Assume that we have a
corpus $W$ of $m$ documents, and a document d is a vector of $n$
words, where each word is selected from a vocabulary of $|V|$ words.
We define document $d=\vec{w}=(w_{1},w_{2},\ldots,w_{n})$, and corpus
$W=(\vec{w_{1}},\vec{w_{2}},\ldots,\vec{w_{m}})$. In unigram model,
the probability of generating document $d$ is:

\begin{equation}
p(\vec{w})=p(w_{1},w_{2},\ldots,w_{n})=p(w_{1})p(w_{2})\cdots p(w_{n})\label{eq:Eq.(1)}
\end{equation}
Considering that the documents are interchangeable with each other,
the probability of generating corpus $W$ is:

\begin{equation}
p(\vec{W})=p(\vec{w_{1}})p(\vec{w_{2}})\cdots p(\vec{w_{m}})\label{eq:Eq.(2)}
\end{equation}
Suppose the word probability of this corpus is $N$, and the probability
of word $i$ in corpus is $n_{i}$. Then written in full, $\vec{n}=(n_{1},n_{2},\ldots,n_{V})$
can be representative as a multinomial distribution over $V$ words
vocabulary:

\begin{equation}
p(\vec{n})=Mult(\vec{n}|\vec{p},N)=\left(\begin{array}{cc}
N\\
\vec{n}
\end{array}\right)\prod_{k=1}^{V}p_{k}^{n_{k}}\label{eq:Eq.(3)}
\end{equation}
where $\vec{p}=(p_{1},p_{2},\ldots,p_{V})$ is the vector of $V$
words probability in vocabulary, and the probability of the corpus
is $p(\vec{W})=p(\vec{w_{1}})p(\vec{w_{2}})\cdots p(\vec{w_{V}})=\prod_{k=1}^{V}p_{k}^{n_{k}}$.

\subsection{LDA Model}

LDA model is a Bayesian hierarchy topic model \cite{Fei05}, generating
topic words for each document with efficiently reduced complexity.
Also, LDA model characterize the possibility that one document may
contain multiple topics, while unigram model only consider the single
topic situation. Instead of calculating probability of word frequency
using continually multiply in unigram model, LDA model maps a document
of $N$ words $d=\vec{w}=(w_{1},w_{2},\ldots,w_{n})$ to $|T|$ latent
topics. As the hierarchy model shows in Figure 1, document-word distribution
has been mapped to document-topic distribution following topic-word
distribution. Therefore, the general expression for word probability
in topic model is:

\begin{equation}
p(w|d)=\sum_{j=1}^{T}p(w|z_{j})\cdot p(z_{j}|d),\label{eq:Eg.(4)}
\end{equation}
where $z_{j}$ is the $j$th topic sampled from a multinomial distribution
of which the prior distribution is a Dirichlet distribution.

\begin{figure}[h]
\centering
\includegraphics[scale=0.5]{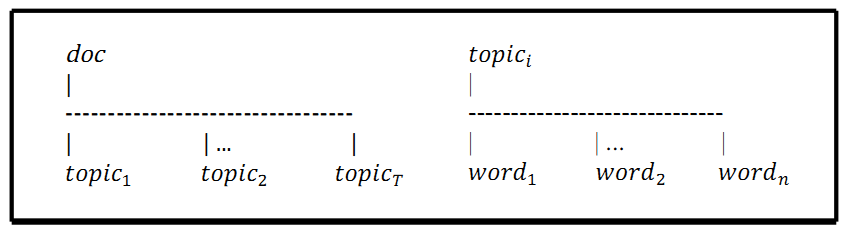}
\caption{Graphic representation of hierarchy topic model.
Each document is mapped to a mixture of $|T|$ topics from document-topic
distribution, and then each one of $n$ words is generated under its
latent topic by sampling from the topic-word distribution.}
\label{fig:one}
\end{figure}


In our project, a document of $n$ words $d=\vec{w}=(w_{1},w_{2},\ldots,w_{n})$
is generated in the following process. Suppose that there are $|T|$
latent topics, then the probability of $i$th word $w_{i}$ in the
given document can be represented in the following mixture:

\begin{equation}
p(w_{i})=\sum_{j=1}^{T}p(w_{i}|z_{i}=j)\cdot p(z_{i}=j),\label{eq:Eq.(5)}
\end{equation}
where $z_{i}$ is the topic to which $i$th word $w_{i}$ assigned,
$p(w_{i}|z_{i}=j)$ represents the probability of word $w_{i}$ assigned
to the $j$th topic, and $\sum_{j=1}^{T}p(z_{i}=j)$ gives the topic
mixture proportion for the current sampled document. Assume that the
corpus is a collection of $|D|$ documents and the vocabulary of this
corpus has $|V|$ unique words. Each document $d$ of $|N_{d}|$ words
is generated according to $|T|$ topics. Let $\phi_{w}^{(z=j)}$ denote
$p(w_{i}|z_{i}=j)$, representing that word $w_{i}$ is sampled from
the multinomial distribution on the $j$th topic $z_{j}$. And let
$\psi_{z=j}^{(d)}$ denote $p(z_{i}=j|d)$, which is a multinomial
distribution from $|T|$ topics for document $d$. Therefore, the
probability of word $w$ in document $d$ is:

\begin{equation}
p(w|d)=\sum_{j=1}^{T}\phi_{w}^{(z=j)}\cdot\psi_{z=j}^{(d)}\label{eq:Eq.(6)}
\end{equation}
In LDA model, $\psi^{(d)}$ sampled from $Dirichlet(\alpha)$ is the
prior distribution of multinomial distribution $\psi_{z=j}^{(d)}$
\cite{Evans11}, and $\phi^{(z)}$ sampled from symmetric $Dirichlet(\chi)$
is the prior distribution of multinomial distribution $\phi_{w}^{(z=j)}$.
Then the multinomial distributions $\phi_{w}^{(z=j)}$ and $\psi_{z=j}^{(d)}$
in LDA model is parameterized as follows:

\begin{equation}
w_{i}|z_{i},\phi^{(z_{i})} Mult(\phi^{(z_{i})}),\quad\phi^{(z_{i})} Dirichlet(\chi)\label{eq:Eq.(7)}
\end{equation}

\begin{equation}
z_{i}|\psi^{(d_{i})} Mult(\psi^{(d_{i})}),\quad\psi^{(d_{i})} Dirichlet(\alpha)\label{eq:Eq.(8)}
\end{equation}
In Eq.(\ref{eq:Eq.(7)}), $\chi$ is a $|T|\times|V|$ matrix, which
is the initial value of word probability sampled from $|T|$ topics.
And in Eq.(\ref{eq:Eq.(8)}), $\alpha=<\alpha_{1},\alpha_{2},\ldots,\alpha_{T}>$
is the initial value of topic probability. $\chi$ and $\alpha$ are
parameters of prior distribution of each multinomial distribution.
We assume both prior distributions to be symmetric Dirichlet distributions.
Therefore, $\chi$ is initialed the same value in the beginning of
sampling every document. Also, is initialed the same value in the
beginning of sampling every document.

Figure 2 shows the Bayesian network \cite{Carey03} of LDA model.
The plates represent repeated sampling. In the left part, the inner
plate represents generating each topic and each word under its topic
repeatedly in a document $d$; the outer plate represents repeated
sampling topic proportion for each of $|D|$ documents in corpus.
And the right plate repeatedly samples $|T|$ parameters for $Mult(\phi^{(z_{i})})$.

\begin{figure}[h]
\centering
\includegraphics[scale=0.5]{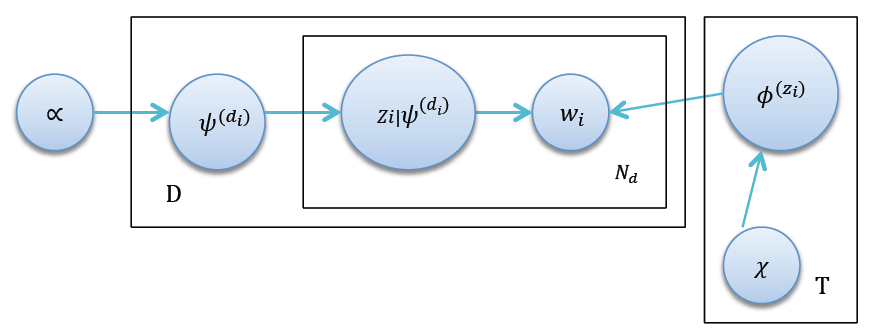}
\caption{Bayesian network of LDA model.}
\label{fig:two}
\end{figure}

\section{Gibbs Sampling}

To estimate parameters of LDA model, Lada Adamic and Eytan Adar use
Expectation Maximization (EM) algorithm as the inference
strategy \cite{McCallum99}. In our project, Gibbs sampling is the choice. Considering
the posterior distribution $p(w|z)$, Gibbs sampling is a simple strategy
to estimate $\phi$ and $\psi$. As a simple case of Markov chain
Monte Carlo (MCMC) algorithm, Gibbs sampling aims at constructing
a Markov chain converging to the target distribution on $z$, and
selecting samples approximating the inferred distribution. The sampling
method begins with initialing the value of vector $z$. Then it repeatedly
samples the $z_{i}$ from the conditional probability $p(z_{i}=j|z_{-i},w_{i})$
and transfers to the next state of Markov chain by updating the probability
function using the newly sampled $z_{i}$. In our project, the probability
function of Gibbs sampling is:

\begin{equation}
P(z_{i}=j|z_{-i},v_{i})=\frac{\frac{n_{-i,j}^{(v_{i})}+\chi}{n_{-i,j}^{(.)}+V\chi}\cdot\frac{n_{-i,j}^{(d_{i})}+\alpha}{n_{-i,.}^{(d_{i})}+T\alpha}}{\sum_{j=1}^{T}\frac{n_{-i,j}^{(v_{i})}+\chi}{n_{-i,j}^{(.)}+V\chi}\cdot\frac{n_{-i,j}^{(d_{i})}+\alpha}{n_{-i,.}^{(d_{i})}+T\alpha}},\label{eq:Eq.(9)}
\end{equation}
where $z_{i}=j$ stands for the assignment of word $v_{i}$, the $i$th
word in a document, to topic $j$; and $z_{-i}$ represents all $z_{k}\,(k\neq i)$
assignments. $n_{-i,j}^{(v_{i})}$ is the number of times $v_{i}$
assigned to topic $j$; $n_{-i,j}^{(.)}$ is the number of words in
vocabulary assigned to topic $j$; $n_{-i,j}^{(d_{i})}$ is the number
of words in document $d_{i}$ assigned to topic $j$; all numbers
of words do not include the current assignment $z_{i}=j$. 

The detailed sampling process is as follows:
\begin{enumerate}
\item For $i=1$ to $N$, where $N$ is the number of word in the current
document, iteratively initial $z_{i}$ as one random integer between
$1$ to $T$. This $N$-sized $Z$ vector is the initial state of
this Markov chain.
\item For $i=1$ to $N$, transfer to next state of this Markov chain by
iteratively assigning word $v_{i}$ to its topic using Eq.(\ref{eq:Eq.(9)}).
\item Run step 2 for $b$ iterations until it reaches the convergent state.
For $i=1$ to $N$, the current value of $z_{i}$ is selected as a
sample. The value of $b$ is called Burn-in period in Gibbs sampling.
\end{enumerate}

Eq.(\ref{eq:Eq.(9)}) is the function calculating the posterior distribution
over word distribution in each document. Therefore, we can derive
the conditional probability function estimating $\phi$ and $\psi$
for every unique word $w$ in a document by removing the word tag
$i$ in Eq.(\ref{eq:Eq.(9)}):

\begin{equation}
\tilde{\phi}_{w}^{(z=j)}=\frac{n_{j}^{(v)}+\chi}{n_{j}^{(.)}+V\chi},\quad\tilde{\psi}_{z=j}^{(d)}=\frac{n_{j}^{(d)}+\alpha}{n_{.}^{(d)}+T\alpha},\label{eq:Eq.(10)}
\end{equation}

where $n_{j}^{(v)}$ is the number of times $v$ assigned to topic
$j$; $n_{j}^{(.)}$ is the number of words in vocabulary assigned
to topic $j$; $n_{j}^{(d)}$ is the number of words in document $d$
assigned to topic $j$; $n_{.}^{(d)}$ is the number of all words
in document $d$ assigned to its topic.

\section{Data Preprocessing}

In this section, we discuss the operations of data preprocessing for
LDA model. \textit{Datagreening.com} server, Python packages and BashReduce
tool implemented in our project will be introduced in the following.

\subsection{Datagreening.com}

The input data for analyzing in LDA model are emails sent to our locally
registered account at \textit{Obama.com}. In our project, these emails
are received through \textit{Datagreening.com}, which is an email
server developed by our research group. This server provides email
service with clean energy and collects research data of carbon footprint
in the meantime. Also, capturing the energy cost in datacenters of
popular email providers, this greening server helps further research
in performance of cloud computing.

\subsection{Python packages}

The downloaded emails are processed to documents consisting of unique
terms using Python NumPy package \cite{NumPy13} and Natural Language Toolkit
(NLTK) package \cite{NLTK13}. NumPy is the fundamental package for scientific
computing with Python. We install the NumPy package to offer the proper
back environment for NLTK packages and include sorting algorithm functions
for Python code. NLTK is a Python platform for text processing. Some
of the NLTK packages are installed for raw content tokenizing, word
stemming and stop words removing in our project. Following is the
list of packages we installed:
\begin{enumerate}
\item banktree\_tag package;
\item stop\_word package;
\item wn package. 
\end{enumerate}

\subsection{BashReduce tool}

Now we format the unique terms learned by the above Python program
as:

\[ N~word\_1:count\_1,word\_2:count\_2,...,word\_n:count\_n \]
where \textit{N} is the number of unique terms in the current document;
\textit{word\_i} is an integer that indexes this term in the vocabulary.
BashReduce tool is used here to calculate the word count. 

BashReduce is a parallel computing tool applying online MapReduce
\cite{Dean08} model to bash environment. According to BashReduce operation
instruction, we begin with specifying the host list to be \textit{bash.xi.0}
and \textit{bash.xi.1} using BashReduce option \textit{'br –h'}. Thus,
we have two instances for parallel calculating word count. Then we
write Python programs \textit{map.py} and \textit{reduce.py}. Program
\textit{map.py} maps each word in word set to pattern (word,1); while
\textit{reduce.py} receives such patterns and sum the same patterns
to generate the (word, count) result and implements. We use BashReduce
options \textit{'-m'} and \textit{'-r'} to run \textit{map.py} and
\textit{reduce.py} respectively. Note that, both input and output
paths for \textit{map.py} and \textit{reduce.py} need to be defined
by user. Running on two instances implemented BashReduce, it costs
3'41'' to finish the word count program, ignoring the network latency.
Comparing with the time cost 5'18'' when calculating locally, it
achieve almost 30\% speedup. 

\section{Experimental Result}

In our results, we use emails sent to our locally registered account
by \textit{Obama.com}. There are total 58 emails in this account with
a vocabulary size of $|V|=1118$ words. To use Gibbs sampling more
efficiently, we first fix the value of burn-in period $b$ using the
data learned in section 4. Then we show the experimental result of
top 15 words generated under 5 latent topics. And we define precision
rate to evaluate the predictive power.

\begin{figure}[h]
\centering
\includegraphics[scale=0.7]{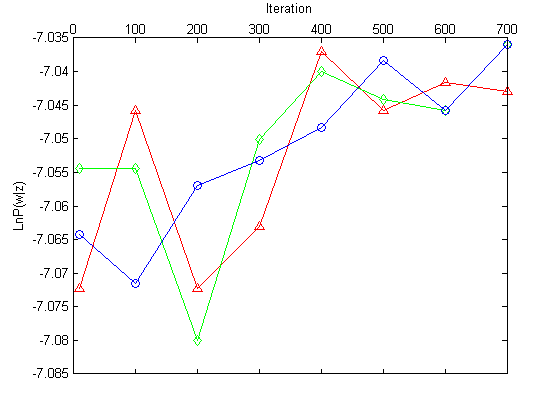}
\caption{Bayesian network of LDA model.}
\label{fig:three}
\end{figure}

\subsection{Burn-in period choice}

We define the number of topics $|T|=300$ and use 3 distinct values
for initialization. Then the convergent sample result $\ln P(w|z)$
will be obtained when choosing the appropriate iteration number. 

Figure 3 shows the convergent process for Gibbs sampling. Starting
at 3 distinct values, the sample result tends to gather and reach
a constant value independent with all three initial values after 500
iterations. 

\subsection{Experimental Result}

In our experiment, parameters $\chi$ and $\alpha$ are assigned values
0.01 and $50/T$ respectively. Latent topics are selected after 500
iterations in Gibbs sampling. And words are generated under 300 topics.
Table 1 give the example of top 15 words generated under 4 topics.

%

\begin{table}[h]
\tbl{Example of top 15 words generated under 5 topics.\label{tab:one}}{%
\begin{tabular}{|c|c|c|c|c|}
\hline
Topic1 & Topic2 & Topic3 & Topic4 & Topic5\\\hline
action & winner & ohio & display & president\\\hline
organize & enter & party & color & obama\\\hline
email & guest & help & none & contact\\\hline
make & action& need & medium & supporter \\\hline
take & organize & governor & screen & committee \\\hline
get & nbsp & state & input & email \\\hline
ofa & washington & authorize & block & democrat \\\hline
people & receive & friend & auto & let \\\hline
health & contribution & make & http & washington \\\hline
fight & entry & kasich & label & work \\\hline
send & email & campaign & leave & know \\\hline
care & state & pay & nbsp & candidate \\\hline
Box & ticke & republican & see & support \\\hline
friend & prize & voter & table & country \\\hline
address & resident& Work & Arial & stay \\\hline
\end{tabular}}
\begin{tabnote}
\tabnoteentry{$^a$}{Words in each topic list is able to help reveal what corresponding
topic it can be. According to the table, a email from \textit{Obama.com}
is likely consisting of health care (Topic1), new contributions (Topic2),
local news (Topic3), and president information (Topic4).}
\end{tabnote}
\end{table}

\subsection{Result Analysis }

To compare the LDA model applied in our project with other models,
we intuitively define a word list of size $|C|=15$, and each word
in the target list works as an identifier to represent the information
of this corpus.

\textit{correct list = \{obama, ohio, health, washington, governor,campaign, republican, president, party, supporter, state, committee,democrat, voter, work\}}\\
And the target word list are defined as part of the correct list. 

\subsubsection{Evaluation measure}
We use precision as a measure to evaluate the experimental results.
The definition of precision is defined as follow:

\begin{equation}
precision=\frac{n_{correct}}{n_{total}}\label{eq:Eq.(11)}
\end{equation}

\begin{figure}[htbp]
\centering
\includegraphics[scale=0.7]{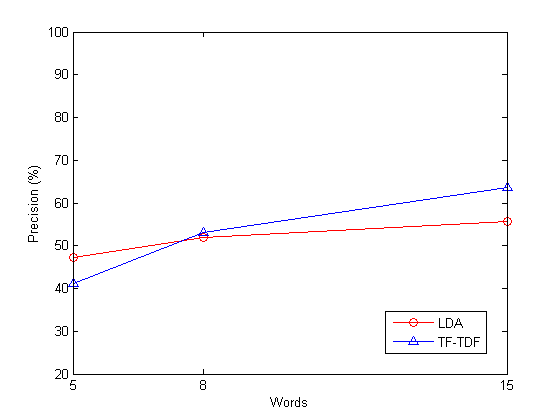}
\caption{Precision rate of LDA model and TF-IDF model when
$|TW|=5$. Precision rate of both model grows when $|TG|$ increases,
for the reason that larger $|TG|$ may lead to more matches. Considering
the limited $|TW|$, the precision rate of LDA model falls below TF-IDF
model when $|TG|>8$.}
\label{fig:four}
\end{figure}

\begin{figure}[htbp]
\centering
\includegraphics[scale=0.7]{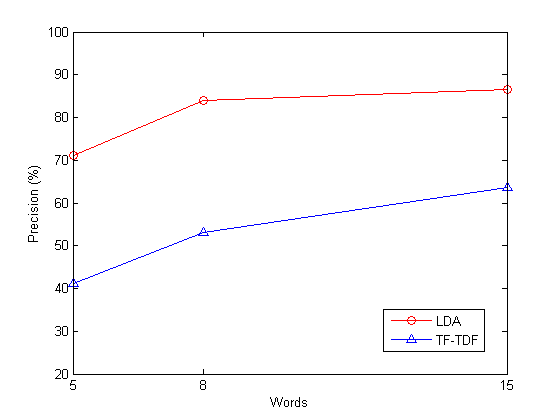}
\caption{Precision rate of LDA model and TF-IDF model when
$|TW|=10$. Precision rate of both model grows when increases $|TG|$.
Higher precision rate of LDA model than TF-IDF model illuminates that
the predictive power of LDA model is stronger than TF-IDF model, when
$|TW|$ is large enough.}
\label{fig:five}
\end{figure}

\begin{figure}[htbp]
\centering
\includegraphics[scale=0.7]{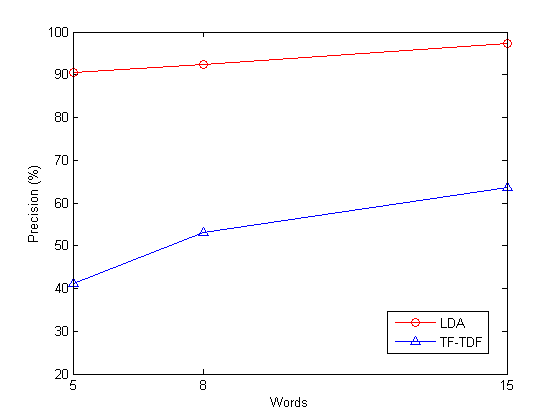}
\caption{Precision rate of LDA model and TF-IDF model when
$|TW|=15$. Precision rate of both model grows when increases $|TG|$.
And when $|TW|=15$, the precision rate of LDA model is almost 53.96\%
higher than TF-IDF model.}
\label{fig:six}
\end{figure}

For each document with $|T|$ topic word list, we compare the top
$k$ words of each topic words list with the words in current target
word list. If any match is captured, then we mark this document as
'correct'. $n_{correct}$ in Eq.(\ref{eq:Eq.(11)}) represents the
number of 'correct' documents in this corpus, and $n_{total}$ stands
for the total number of documents in this corpus. 

\subsubsection{Comparison model}

We choose TF-IDF model as a comparison.TF-IDF, short for Term Frequency-Inverse Document Frequency model,
a numerical statistic representing the relation between the key word
and the document, is one of the major weight factors in text mining.
It is often used for text classification by ranking a document's relevance
given a user query. Assume that $D$ is the total number of dicuments
in this corpus; $f(t,d)$ is the raw frequency of a term in a document,
then the expression of this model is:

\begin{equation}
tfidf(t,d,D)=tf(t,d)\times idf(t,D)\label{eq:Eq.(12)}
\end{equation}

where $idf(t,D)$ is defined as:

\begin{equation}
idf(t,D)=\log\frac{|D|}{|\{d\in D:t\in d\}|}\label{eq:Eq.(13)}
\end{equation}

Therefore, a high weight in TF-IDF is reached by a high term frequency
and a low document frequency of the term in the whole collection of
documents; the weights hence tend to filter out common terms and select
the terms with lower probability to be the words that distinguish
documents.

\subsubsection{Predictive power}

Let $|TG|$ be denoted as the size of target word list, and $|TW|$
as the size of topic word list. We evaluate the predictive power by
comparing precision rate for LDA model and TF-IDF model in cases with
different $|TW$|. The following Figure 4, Figure 5 and Figure 6 show
the comparison of LDA model and TF-IDF model where \textit{x}-coordinate
represents different values of $|TG|$ and \textit{y}-coordinate represents
precision rate.







\section{Conclusion}

In this project, we apply LDA model to analyze and vilsualize
topic words of emails from \textit{Obama.com} to accounts registered
in Columbus, Ohio. We use Gibbs sampling to estimate topic and word
distribution. Carbon-free server \textit{Datagreening.com}, pareleling
tool BashReduce and Python packages are used in data preprocessing.
The experimental result shows the MapReduce method in our project
efficiently reduces the time cost of word count program and the predictive
power of LDA model applied in our project is obviously stronger than
TF-IDF model.

Every new document adding to this corpus lead to another
sampling process for it, highly increasing the time cost. Considering
this drawbask, possible future work of our project is likely to be
sampling only new added words instead of a new added document. 

\bibliographystyle{ACM-Reference-Format-Journals}
	\bibliography{LDA}
	
\end{document}